\documentclass[10pt,conference]{IEEEtran}
\usepackage{fixltx2e}
\usepackage{cite}
\usepackage{url}
\usepackage{mathrsfs}
\usepackage{color}
\usepackage{float}
\usepackage[caption=false]{subfig}

\ifCLASSINFOpdf
   \usepackage[pdftex]{graphicx}
   \graphicspath{{Figs/}}
   \DeclareGraphicsExtensions{.pdf,.jpeg,.png}
\else
\fi
\usepackage{balance}
\usepackage[cmex10]{amsmath}
\usepackage{amsmath}
\usepackage{amssymb}
\usepackage{amsthm}
\usepackage{amsfonts}
\usepackage{bm}
\usepackage{xfrac}
\usepackage{empheq}
\usepackage[normalem]{ulem} % either use this (simple) or
\usepackage{soul} % use this (many fancier options)
\usepackage{mathtools}

\newtheorem{theorem}{Theorem}

\newtheorem{example}{Example}

\newtheorem*{observ}{Observation}

\newtheorem{corollary}{Corollary}
\newtheorem{remark}{Remark}
\theoremstyle{definition}

\usepackage[a4paper,bindingoffset=0.2in,%
left=1in,right=1in,top=1in,bottom=1in,%
footskip=.25in]{geometry}
\graphicspath{{figs/}}

\begin{document}
	\newgeometry{left=0.7in,right=0.7in,top=.5in,bottom=1in}
	\title{Multi-Task Private Semantic Communication}
\vspace{-9mm}
\author{
		\IEEEauthorblockN{Amirreza Zamani, Sajad Daei, Tobias J. Oechtering, Mikael Skoglund \vspace*{0.5em}
			\IEEEauthorblockA{\\
				Division of Information Science and Engineering, KTH Royal Institute of Technology \\
				%$^\ddagger$Dept. of Electrical and Electronic Engineering, Imperial College London\\
				Email: \protect amizam@kth.se, sajado@kth.se, oech@kth.se, skoglund@kth.se }}%\vspace*{-2.1em}
		}
	\maketitle

\begin{abstract}
	We study a multi-task private semantic communication problem, in which an encoder has access to an information source arbitrarily correlated with some latent private data. A user has $L$ tasks with priorities. The encoder designs a message to be revealed which is called the semantic of the information source. %Here, $h(X)$ denotes the goal or task of the receiver and $f(X)$ corresponds to the semantic. 
	Due to the privacy constraints the semantic can not be disclosed directly and the encoder adds noise to produce disclosed data. The goal is to design the disclosed data that maximizes the weighted sum of the utilities achieved by the user while satisfying a privacy constraint on the private data.
	%The design of a statistical signal processing privacy problem is studied where the private data is assumed to be observable. 
	%We study privacy mechanism design problems with information theory perspective, where 
	%two scenarios are considered.
	%In this work, 
	%an agent observes useful data $Y$, which is correlated with private data $X$, and wants to disclose the useful information to a user. A statistical privacy mechanism is employed to generate data $U$ based on $(X,Y)$ that maximizes the revealed information about $Y$ while satisfying a privacy criterion. % where bounded leakages are considered, i.e., $0\leq I(X;U)\leq\epsilon$. In this case, $X-Y-U$ forms a Markov chain. However, 
	%In the second scenario, the agent has additionally access to the private data. % maximizes the revealed information about $Y$ while satisfying the privacy constraint. 
	
	 In this work, we first consider a single-task scenario and design the added noise utilizing various methods including the extended versions of the Functional Representation Lemma, Strong Functional Representation Lemma, and separation technique. We then study the multi-task scenario and derive a simple design of the source semantics. We show that in the multi-task scenario the main problem can be divided into multiple parallel single-task problems.
\end{abstract}
\section{Introduction}
Semantic communication involves transmitting a modified version of the original information source with reduced dimensionality to a receiver, whose objective is to extract information for a specific goal or task \cite{sherdeniz}. %which is a lower-dimensional subset of the original information. 
%On the other hand, the goal is to design privacy mechanisms with optimal privacy and utility trade-off.
Semantic communication considers not just the literal message, but also its context, connotations, and nuances, aiming to prevent ambiguity and misunderstandings by aligning it with the recipient's perception, knowledge, expectations, and cultural context \cite{feist2022significance,sherdeniz}. Semantic communications have received significant research attention in recent years as a method to reduce the data load on 6G and future networks by transmitting only the semantically relevant information to the receiver.

Another important dimension of emerging communication systems is data privacy. With increasing reliance on data for a wide range of applications, it is essential for users to only reveal non-sensitive information to receivers. In that sense, semantic communication already meets
some privacy considerations by preventing the sharing of
sensitive information. On the other hand, various tasks may
require sending information correlated with some private
attributes, which may be hard to identify and protect.  

Related works on the semantic communications and information theoretic approach to the privacy can be found in
\cite{sherdeniz,feist2022significance,strinati20216g,shao2022theory,borz,khodam,Khodam22,kostala,shah, makhdoumi, dwork1, shahab,liu2020robust,sankar, oech, Total, sankar2, deniz4, asoodeh3, Calmon1,  nekouei2, issa,oof}. 
 
 %Excluding the privacy concerns, relevant scenarios have been studied in \cite{strinati20216g} and \cite{shao2022theory}.
%In both \cite{yamamoto} and \cite{sankar}, the privacy-utility trade-off is considered using expected distortion as a measure of utility and equivocation as measure of privacy. 
%In \cite{Calmon2}, fundamental limits of the privacy utility trade-off measuring the leakage using estimation-theoretic guarantees are studied.

%In \cite{yamamoto}, a source coding problem with secrecy is studied.
%Privacy-utility trade-offs considering equivocation as measure of privacy and expected distortion as a measure of utility are studied in \cite{sankar}.
In \cite{borz}, the problem of privacy-utility trade-off considering mutual information both as measures of privacy and utility is studied. Under perfect privacy assumption, it has been shown that the privacy mechanism design problem can be reduced to linear programming. %This work has been extended in \cite{gun} considering the privacy utility trade-off with a rate constraint for the disclosed data.
%Moreover, in \cite{borz}, it has been shown that information can be only revealed if $P_{X|Y}$ is not invertible. 
In \cite{khodam}, privacy mechanisms with a per letter privacy criterion considering an invertible leakage matrix have been designed allowing a small leakage. This result is generalized to a non-invertible leakage matrix in \cite{Khodam22}.
%In \cite{borz}, the problem of maximizing utility, i.e., $I(U;Y)$, under the leakage constraint $I(U;X)\leq \epsilon$ and Markov chain $X-Y-U$ is studied and it is shown that under perfect privacy assumption, i.e., $\epsilon=0$, the privacy mechanism design problem can be reduced to standard linear programming.
\begin{figure}[]
	\centering
	\includegraphics[scale = .5]{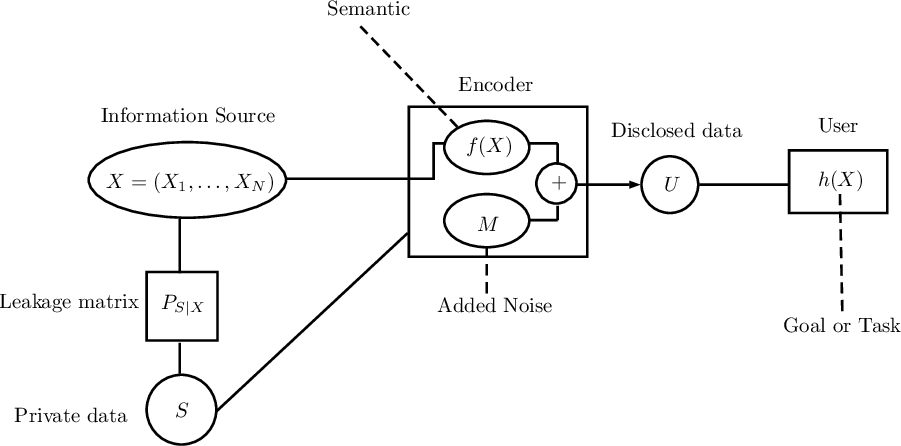}
	\caption{Single-task private semantic communication model. The goal is to design disclosed data $U$ such that it keeps as much information as possible about the task $h(X)$ while satisfying a certain privacy constraint.}
	\label{semsysch112}
\end{figure}
%Here we consider the problem studied in  \cite{borz}, \cite{kostala}, \cite{shahab}, and \cite{king1}. 
In \cite{kostala}, \emph{secrecy by design} problem is studied under the perfect secrecy assumption. Bounds on secure decomposition have been derived using the Functional Representation Lemma. %and new bounds on privacy-utility trade-off are derived. %The bounds are tight when the private data is a deterministic function of the useful data. 
In \cite{shah}, the privacy problems considered in \cite{kostala} are generalized by relaxing the perfect secrecy constraint and allowing some leakage. In \cite{9457633}, fundamental limits of private data disclosure are studied, where the goal is to minimize leakage under utility constraints with non-specific tasks.  %More specifically, we considered bounded mutual information, i.e., $I(U;X)\leq \epsilon$ for privacy leakage constraint. 
%Moreover, the bounds obtained in \cite{king1} have been tightened in \cite{zamani2023new} by using \emph{separation technique}.
%Furthermore, in the special case of perfect privacy we derived a new upper bound for the perfect privacy function and it has been shown that this new bound generalizes the bound in \cite{kostala}. Moreover, it has been shown that the bound is tight when $|\mathcal{X}|=2$.

In the present work, we utilize concepts from the privacy mechanism design outlined in \cite{shah} to introduce a novel multi-task private semantic communication framework. The proposed scheme offers a mathematical approach to design a goal-oriented privacy mechanism. This mechanism not only facilitates the receiver in achieving its goal but also guarantees the privacy of the sensitive data from the recipient.
%and combine them with semantic communication as discussed in \cite{strinati20216g} and \cite{shao2022theory}. 
To design the disclosed data we consider different methods. Extended versions of the Functional Representation Lemma and the Strong Functional Representation Lemma, as well as a separation technique are used to address the \emph{multi-task private semantic communication} problem.
%To this end, we use extended versions of the Functional Representation Lemma and the Strong Functional Representation Lemma and combine them with a simple observation. 
The Functional Representation Lemma (FRL) \cite[Lemma 1]{kostala} and the Strong Functional Representation Lemma (SFRL) \cite[Theorem 1]{kosnane} are constructive lemmas that are valuable for the privacy mechanism design.
The \emph{separation technique} corresponds to representing a discrete random variable (RV) by two correlated RVs. % We call this observation \emph{separation technique} since it separates a RV into two RVs. 
Privacy preserving message is obtained by adding noise to the semantic information. %In this work we assume that both semantic and goal are known by the encoder. 
We first study a single-task scenario and provide lower and upper bounds on the privacy-utility trade-off. We then consider a multi-task scenario where the user has different goals with varying priorities. We propose a design that maximizes a weighted linear combination of the utilities achieved by the user while satisfying a privacy constraint. We show that the highly challenging and complex problem can be decomposed into independent single-task problems and a simple privacy mechanism design can be obtained.

\vspace{-1mm}
\section{system model and Problem Formulation} \label{sec:system}
\vspace{-1mm}
\textbf{Single-task scenario:}
Let $P_{SX}=P_{SX_1,\cdots,X_N}$ denote the joint distribution of discrete random variables $X=(X_1,\ldots,X_N)$ and $S$ defined on finite alphabets $\mathcal{X}=\mathcal{X}_1\times\cdots\times\mathcal{X}_N$ and $\cal{S}$. 
Here, $S$ denotes the latent private data, while $X$ is the information source of dimension $N$. 
%We assume that cardinality $|\mathcal{X}|$ is finite and $|\mathcal{Y}|$ is finite or countably infinite.
We represent $P_{SX}$ by a matrix defined on $\mathbb{R}^{|\mathcal{S}|\times|\mathcal{X}|}$. %and %We assume that $X$ and $Y$are defined on spaces that have the same cardinality, i.e., $|\cal{X}|=|\cal{Y}|=\mathcal{K}$. 
%marginal distributions of $S$ and $X$ by vectors $P_S$ and $P_X$ defined on $\mathbb{R}^{|\mathcal{S}|}$ and $\mathbb{R}^{|\mathcal{X}|}$ given by the row and column sums of $P_{SX}$. 
%We assume that each element in vectors $P_X$ and $P_Y$ is non-zero. Furthermore, 
We represent the leakage matrix $P_{S|X}$ by a matrix defined on $\mathbb{R}^{|\mathcal{S}|\times|\cal{X}|}$.
%In this work, RVs $X$ and $Y$ denote the private data and the useful data and $U$ describes the disclosed data. 
\iffalse
As we mentioned earlier, we consider two scenarios.
In the first scenario, we aim to design a privacy mechanism that produces disclosed data $U$ based on $Y$ that maximizes $I(U;Y)$ and satisfies the leakage constraint $I(U;X)\leq \epsilon$. In other words, in the first scenario the Markov chain $X-Y-U$ holds. However, in the second scenario, the goal is to design a privacy mechanism that produces disclosed data $U$ based on the pairs $(X,Y)$ that maximizes $I(U;Y)$ and satisfies the leakage constraint $I(U;X)\leq \epsilon$.\fi
%Therefore,
The semantic of $X$ is defined as a function of $X$ denoted by $f(X):\mathcal{X}_1\times\cdots\times\mathcal{X}_N\rightarrow \mathbb{R}^T$ with dimension $T\leq N$. Furthermore, the goal or task of communication is represented by some other function of $X$, i.e., $h(X):\mathcal{X}_1\times\cdots\times\mathcal{X}_N\rightarrow \mathbb{R}^K$, with dimension $K\leq N$. In most cases $K$ is significantly smaller than $N$ since the semantic is designed based on the goal. In general $f\neq h$, since $f(\cdot)$ needs to be designed in an efficient way and efficiency can be defined based on different parameters.     
%\begin{figure}[]
%	\centering
%	\includegraphics[scale = .5]{semantic}
%	\caption{Proposed approach for dealing with privacy concerns: adding artificial noise to the semantic $f(X)$.}
%	\label{semsysch112}
%\end{figure}
As shown in Fig. \ref{semsysch112}, the noise denoted by a discrete RV $M\in \mathcal{M}$ is added to the goal to produce disclosed data described by RV $U\in \mathcal{U}$. However, it can annihilate the performance by decreasing the utility achieved by the user. Following this approach leads to a privacy-utility trade-off problem. 
 In this model, the leakage is measured by the mutual information between $S$ and $U$. Furthermore, the utility achieved by the user is measured by the mutual information between $U$ and $h(X)$. 
We assume that both the semantic and the goal are known to the encoder and the task is to design the privacy mechanism, i.e., added noise, that achieves the optimal trade-off between the privacy and the utility.

\begin{figure}[]
	\centering
	\includegraphics[scale = .22]{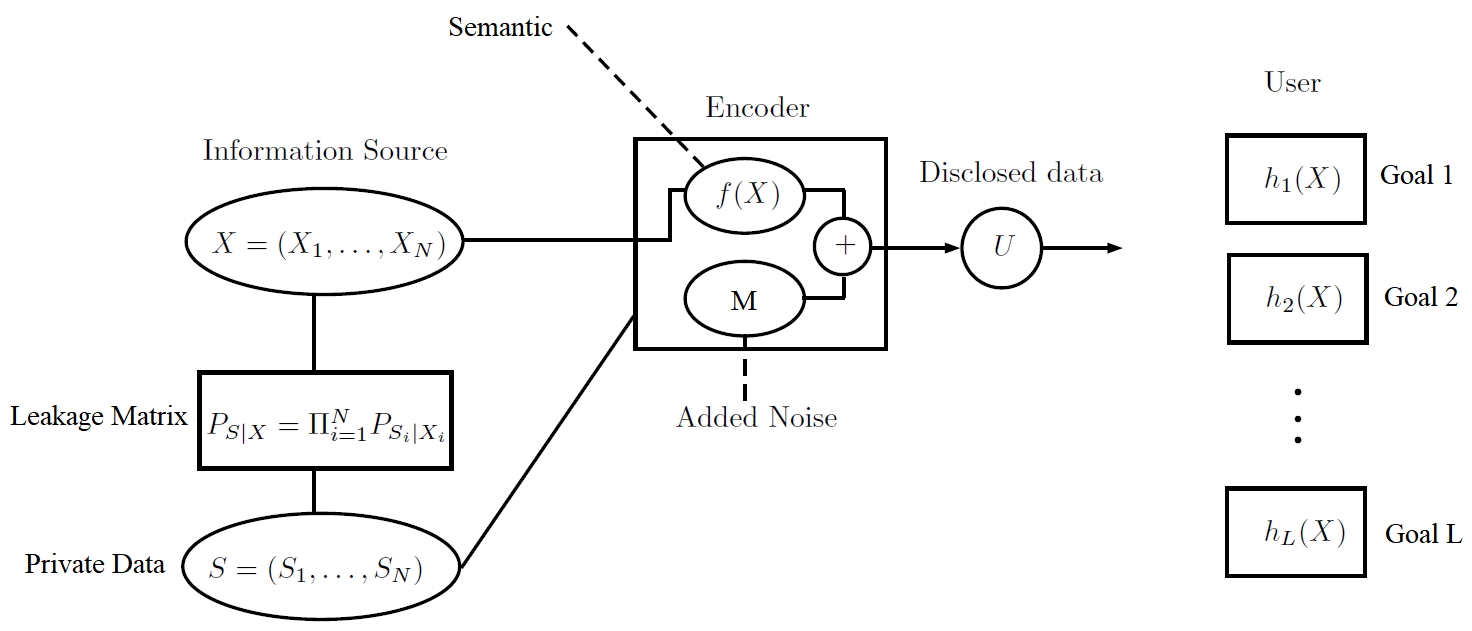}
	\caption{Multi-task private semantic communication model. The goal is to design $U$ that maximizes weighted linear combination of utilities while satisfying a certain privacy constraint.}
	\label{sem}
\end{figure}

%Here, we use mutual information as utility and leakage measures. The privacy mechanism design problem can be stated as follows %we define two privacy preserving functions $g_{\epsilon}(P_{XY})$ and $h_{\epsilon}(P_{XY})$ as follows
The private semantic communication problem can be stated as follows
\vspace{-2mm}
\begin{align}
h_{\epsilon}(P_{S,f(X),h(X)})&=\!\!\!\!\!\!\!\!\sup_{\begin{array}{c} 
	\substack{P_{U|S,f(X),h(X)}: I(U;S)\leq\epsilon,}
	\end{array}}\!\!\!\!\!\!\!\!I(h(X);U),\label{main1ch11}
\end{align} 
where $U=f(X)+M$, $P_{S,f(X),h(X)}$ is the joint distribution of $(S,f(X),h(X))$, and $P_{U|S,f(X),h(X)}$ describes the conditional distribution.
In the following we study the case where $0\leq\epsilon< I(S;h(X))$, otherwise the optimal solution of $h_{\epsilon}(P_{S,f(X),h(X)})$ is $H(h(X))$ achieved by $U=h(X)$, i.e., $M=f(X)-h(X)$.
%\vspace{-2mm}
\begin{remark}
	\normalfont
	In \eqref{main1ch11}, the privacy mechanism design is based on $f(X)$, $h(X)$, and $S$ which are accessible by the encoder. Hence, the optimization is over $P_{U|S,f(X),h(X)}$ instead of $P_{U|S,X}$. In other words, $(S,f(X),h(X))$ is a sufficient statistic and the encoder does not require access to the information source $X$ if it has access to $f(X)$, $h(X)$, and $S$. 
\end{remark}
%\vspace{-3mm}
	A scenario that motivates our model can be stated as follows.
	Assume that the information source is not accessible directly to the encoder. Moreover, there exists a third party that designs $f(X)$ based on the task or goal $h(X)$ and shares it with the encoder. Since $f(X)$ is correlated with $S$, it can not be revealed directly. Thus, the encoder designs $U$ based on $f(X)$, $h(X)$ and $S$.\\
\textbf{Multi-task scenario:}
In this part, as shown in Fig. \ref{sem}, let $X$ and $S$ consist of $N$ independent RVs denoted by $(X_1,\ldots,X_N)$ and $(S_1,\ldots,S_N)$ where $X_i$ and $S_i$ are arbitrarily correlated. Hence, the joint distribution of $X$ and $S$ can be rewritten as $P_{SX}=\prod_{i=1}^N P_{S_i,X_i}(s_i,x_i)$. We extend the problem defined in \eqref{main1ch11} by considering $L$ tasks where each is a function of $X$. Here, $h_{i}(X):\mathcal{X}_1\times\ldots\mathcal{X}_N\rightarrow \mathbb{R}^{K_i}$, with dimension $K_i\leq N$, denotes task $i$. We assume that for each $i$, $h_i(X)=A_iX^T$, where $A_i\in\mathbb{R}^{N\times N}$ is a matrix with at most one non-zero element in each row and column that equals one. In other words, $h_i(X)$ represents a subset of the information source $X=(X_1,\ldots,X_N)$. Based on $L$ tasks the encoder designs the semantic $f(X)$ and adds noise $M$ to it to produce $U$. For each task the user achieves a utility which is measured by the mutual information between the disclosed data and task $i$, i.e., $I(U;h_i(X))$. The goal is to design the noise $M$ such that $U$ maximizes a weighted linear combination of the utilities while satisfying a privacy leakage constraint. Similarly, the privacy leakage is measured by $I(U;S)$. The multi-task private semantic communication problem can be stated as follows %we define two privacy preserving functions $g_{\epsilon}(P_{XY})$ and $h_{\epsilon}(P_{XY})$ as follows
\vspace{-3mm}
\begin{align}
\!\!\!\!h_{\epsilon}^L(P_{S,f(X),\bm{h}(X)})=\!\!\!\!\!\!\!\!\!\!\!\!\!\!\!\!\!\!\!\!\!\!\!\!\sup_{\begin{array}{c} 
	\substack{P_{M|S,f(X),\bm{h}(X)}:I(S;U)\leq\epsilon}
	\end{array}}\!\!\!\sum_{i=1}^L \lambda_iI(h_i(X);U),\label{main1}
\end{align} 
where $U=f(X)+M$, $\bm{h}(X)=\{h_i(X)\}_{i=1}^{L}$, and for $i\in\{1,..,L\}$, $\lambda_i\geq 0$ are fixed.
In the following we study the case where $0\leq\epsilon< I(S;X)$, otherwise the optimal solution of $h_{\epsilon}^L(P_{XY})$ is $\sum_{i=1}^{K} \lambda_i H(h_i(X))$ achieved by $U=X$.
	The weights $\lambda_1,..,\lambda_K$ correspond to the different priorities of the tasks. For instance, if task $i$ is more important than task $j$ we let $\lambda_i\geq \lambda_j$. %be larger than $\lambda_j$.  
\begin{remark}
	In contrast with \cite{liu2020robust}, here, we assume that $X_i$ and $S_i$ are arbitrarily correlated generalizing the assumption that $S_i$ is a deterministic function of $X_i$. %is too restrictive and here we generalize it.
\end{remark}
 \section{Main Results}\label{sec:resul}
 \vspace{-1mm}
 \textbf{Single-task scenario:}
 %In this part, we first present a simple observation which we call ``separation technique''.
 In this part, we provide lower and upper bounds for the privacy problems as defined in \eqref{main1ch11}. We study the tightness of the bounds in special cases and compare them in examples.
%\begin{remark}
%	The representation obtained by the separation technique is not unique. For instance, let $|\mathcal{X}|=16$. In this case, $|\mathcal{X}_1|=4$, $|\mathcal{X}_2|=4$ or $|\mathcal{X}_1|=8$, $|\mathcal{X}_2|=2$.
%\end{remark}
To do so, in Appendix A, we present a simple observation which we call the \emph{separation technique}.
In the following results let $\mathcal{K}_S$ be all possible representations of $S$ using separation technique where $S=(S_1,S_2)$. In other words we have $\mathcal{K}_S=\{(S_1,S_2):S=(S_1,S_2), \ 2\leq|\mathcal{S}_1|<|\mathcal{S}|,\ 2\leq|\mathcal{S}_2|<|\mathcal{S}|\}$. We emphasize that any representation in $\mathcal{K}_S$ corresponds to a non-trivial representation of $S$. For instance, a trivial case is to let $S_1=S$ and $S_2$ be a constant.

Before stating the first theorem we derive an expression for $I(X;U)$. For any correlated random variables $S$, $X$, and $U$, we have
\begin{align}
I(X;U)&=I(S,X;U)-I(S;U|X),\nonumber\\&=I(S;U)\!+\!H(X|S)\!-\!H(X|U,S)\!-\!I(S;U|X).\label{key}
\end{align}
%As argued in \cite{kostala}, \eqref{key} is an important observation to find lower and upper bounds for $h_{\epsilon}(P_{XY})$. 
Next, we derive lower and upper bounds on $h_{\epsilon}(P_{S,f(X),h(X)})$. For lower bounds we use EFRL \cite[Lemma 4]{shah}, ESFRL \cite[Lemma 5]{shah}, and the separation technique.
For simplicity the remaining results are derived under the assumption $f(\cdot):\mathcal{X}_1\times\ldots\mathcal{X}_N\rightarrow \mathbb{R}$, i.e., $T=1$. 
\begin{theorem}\label{semanj}
	For any $0\leq \epsilon< I(S;h(X))$ and joint distribution $P_{S,f(X),h(X)}$, we have
	\begin{align}\label{th2ch11}
	\!\!\!\max_{i\in\{1,\ldots,4\}}\{L_{h}^{i}(\epsilon)\}\!\leq\! h_{\epsilon}(P_{S,f(X),h(X)})\!\leq H(h(X)|S)+\epsilon,
	\end{align}
	where
	\vspace{-2mm}
	\begin{align*}
	L_{h}^{1}(\epsilon) &= H(h(X)|S)-H(S|h(X))+\epsilon\\&=H(h(X))-H(S)+\epsilon,\\
	L_{h}^{2}(\epsilon) &= H(h(X)|S)-\alpha H(S|h(X))+\epsilon\\&-(1-\alpha)\left( \log(I(S;h(X))+1)+4 \right),\\
	L_{h}^{3}(\epsilon) &= H(h(X)|S)\!+\!\epsilon\!-\!\left( \log(I(S;h(X))+1)+4 \right)\\&-\min_{(S_1,S_2)\in \mathcal{K}_S} \{\alpha_2 H(S_2|h(X))\}, \\
	L_{h}^{4}(\epsilon) &=H(h(X)|S)+\epsilon-\left( \log(I(S;h(X))\!+\!1)\!+\!4 \right)\\-\!\!\min_{(S_1,S_2)\in \mathcal{K}_S}\!\!\!&\{\alpha_2 \left(H(S|h(X)-\log(I(S;h(X))\!+\!1)\!-\!4 )\right) \},
	\end{align*}
	with $\alpha=\frac{\epsilon}{H(S)}$ and $\alpha_2=\frac{\epsilon}{H(S_2)}$ for any representation $S=(S_1,S_2)$.
	The lower bound is tight if $H(S|h(X))=0$, i.e., $S$ is a deterministic function of $h(X)$. Furthermore, if the lower bound $L_{h}^{1}(\epsilon)$ is tight then we have $H(S|h(X))=0$.
\end{theorem}
%\begin{proof}
\textbf{Proof:}
	Using \eqref{key} we have
	$
	I(f(X)+M;h(X))=I(f(X)+M;S)+H(h(X)|S)-I(f(X)+M;S|h(X))-H(h(X)|S,f(X)+M),
	$
	which results in 
	$
	I(f(X)+N;h(X))\leq \epsilon+H(h(X)|S).
	$
	For deriving the lower bounds $L_{h}^{1}(\epsilon)$ and $L_{h}^{2}(\epsilon)$ we use EFRL and ESFRL using $S\leftarrow X$ and $h(X)\leftarrow Y$. 
	Let $\bar{U}$ and $\tilde{U}$ be the output of the EFRL and ESFRL. Using the same arguments as in \cite[Theorem 2]{shah} we have
	$
	I(\bar{U};h(X))\geq L_{h}^{1}(\epsilon)$, $
	I(\tilde{U};h(X))\geq L_{h}^{2}(\epsilon)$, and $
	I(\bar{U};S)=I(\tilde{U};S)=\epsilon.
	$
	To achieve $L_{h}^{1}(\epsilon)$ let $M=\bar{U}-f(X)$ and to attain $L_{h}^{2}(\epsilon)$ let $M=\tilde{U}-f(X)$.
	%The lower bound $L_3^{\epsilon}$ can be derived by using \cite[Remark~2]{shahab}, since we have $h_{\epsilon}(P_{XY})\geq g_{\epsilon}(P_{XY})\geq L_3^{\epsilon}$, where $g_{\epsilon}(P_{XY})=\sup_{\begin{array}{c} 
	%	\substack{P_{U|Y,X}: I(U;X)\leq\epsilon,\\ X-Y-U}
	%	\end{array}}I(Y;U)$. The upper bound $U_1^{\epsilon}$ and lower bounds $L_1^{\epsilon}$ and $L_2^{\epsilon}$ have been derived in \cite[Theorem 2]{king1}. The lower bound $L_1^{\epsilon}$ is attained by using \eqref{key} and EFRL. Similarly, the lower bound $L_2^{\epsilon}$ is attained by \eqref{key} and ESFRL, for more detail see \cite[Theorem 2]{king1}. Moreover, the results about tightness have been proved in \cite[Theorem 2]{king1}. It is sufficient to obtain $L_4^{\epsilon}$ and $L_5^{\epsilon}$. The complete proof for obtaining $L_4^{\epsilon}$ and $L_5^{\epsilon}$ is provided in Appendix A. As we mentioned earlier to achieve $L_1^{\epsilon}$ or $L_2^{\epsilon}$ we use EFRL or ESFRL. 
	The main idea for constructing a RV $U$ that satisfies EFRL or ESFRL constraints is to add a randomized response to the output of FRL or SFRL. The randomization introduced in {\cite{warner} is taken over $S$. To derive $L_{h}^{3}(\epsilon)$ and $L_{h}^{4}(\epsilon)$, let $(S_1,S_2)$ be a possible representation of $S$, i.e., $S=(S_1,S_2)$. The main idea to achieve $L_{h}^{3}(\epsilon)$ and $L_{h}^{4}(\epsilon)$ is to take randomization over $S_2$ instead of $S$. In other words, we add a randomized response which is based on $S_2$ instead of $S$. Considering $L_{h}^{2}(\epsilon)$ and $L_{h}^{4}(\epsilon)$, $\alpha$ corresponds to the probability of randomizing over $S$, however, $\alpha_2$ corresponds to the probability of randomizing over $S_2$ for any representation $S=(S_1,S_2)$. 
		The rest of the proof is provided in Appendix B.
	%\end{proof}
	\vspace{-2mm}
\begin{example}
	Let $X=(\bar{X}_1,\bar{X}_2,\bar{X}_3)$, $h(X)=(\bar{X}_1,\bar{X}_2)$, and $S=\bar{X}_1+\bar{X}_2$, where $\bar{X}_1$, $\bar{X}_2$, and $\bar{X}_3$ are arbitrary correlated. In this case since $S$ is a deterministic function of $h(X)$, by using \eqref{th2ch11} the lower bound $L_{h}^{1}(\epsilon)$ is tight and we have
	$
	h_{\epsilon}(P_{S,f(X),h(X)})=H(h(X)|S)+\epsilon=H(\bar{X}_1,\bar{X}_2|\bar{X}_1+\bar{X}_2)+\epsilon.
	$ 
\end{example}
\subsection*{Comparison}\label{khaye}
\vspace{-1mm}
In this part we study the bounds considering different cases. For simplicity let $S=(S_1,S_2)$ where $S_1$ and $S_2$ are arbitrarily correlated. 
In this case we have
\begin{align*}
&U_1^{\epsilon} = H(h(X)|S_1,S_2)-\epsilon,\\
&L_1^{\epsilon} = H(h(X)|S_1,S_2)-H(S_1,S_2|h(X))+\epsilon,\\
&L_2^{\epsilon} = H(h(X)|S_1,S_2)-\alpha H(S_1,S_2|Y)+\epsilon\\&\ -(1-\alpha)\left( \log(I(S_1,S_2;h(X))+1)+4 \right),\\
%L_3^{\epsilon} &= \epsilon\frac{H(Y)}{I(X_1,X_2;Y)}+g_0(P_{XY})\left(1-\frac{\epsilon}{I(X_1,X_2;Y)}\right),\\
&L_{3}^{\epsilon}\geq \bar{L}_{3}^{\epsilon} \triangleq H(h(X)|S_1,S_2)\!+\!\epsilon\\&\!-\!\left( \log(I(S_1,S_2;h(X))\!+\!1)\!+\!4 \right)-\alpha_2 H(S_2|Y), \\
&L_{4}^{\epsilon}\geq\bar{L}_{4}^{\epsilon} \triangleq H(h(X)|S_1,S_2)+\epsilon\\&\!-\!(1\!\!-\!\alpha_2)\!\left( \log(I(S_1,\!S_2;\!h(X)\!)\!+\!1)\!+\!4 \right)\!+\!\alpha_2 H(S_1,S_2|h(X)\!),
\end{align*}
where $\alpha=\frac{\epsilon}{H(S)}$ and $\alpha_2=\frac{\epsilon}{H(S_2)}$.
Note that the lower bounds $\bar{L}_{3}^{\epsilon}$ and $\bar{L}_{4}^{\epsilon}$ are obtained based on the initial representation of $S=(S_1,S_2)$. Therefore, we have $\bar{L}_{3}^{\epsilon}\leq L_{3}^{\epsilon}$ and $\bar{L}_{4}^{\epsilon}\leq L_{4}^{\epsilon}$. Here, we compare the lower bounds $L_{2}^{\epsilon}$, $\bar{L}_{3}^{\epsilon}$, and $\bar{L}_{4}^{\epsilon}$. To do so we consider the following scenarios. Detailed proofs and calculations are provided in supplementary material.\\
\textbf{Scenario 1}: To compare $\bar{L}_{4}^{\epsilon}$ with $L_{2}^{\epsilon}$, let us assume that $H(S_1,S_2|h(X))\leq \log(I(S_1,S_2;h(X))+1)+4$. A simple example can be considering $S_1$ and $S_2$ as binary RVs. In this case we have
\begin{align*}
\bar{L}_{4}^{\epsilon}-L_{2}^{\epsilon}\geq 0.
\end{align*}
%\vspace{-1mm}
\textbf{Scenario 2}: To compare $\bar{L}_{3}^{\epsilon}$ with $L_{2}^{\epsilon}$, let us assume that $S_2$ is a deterministic function of $h(X)$ and $H(S_1|h(X))\geq \log(I(S_1,S_2;h(X))+1)+4$. A simple example is to let $4+H(h(X))\leq H(S_1|h(X))$ which leads to $H(S_1|h(X))\geq \log(I(S_1,S_2;h(X))+1)+4$. In this case we have
%\begin{align*}
%&\bar{L}_{3}^{\epsilon}-L_{2}^{\epsilon}=\\&\frac{\epsilon}{H(S_1,S_2)}\left(H(S_1|h(X))\!-\!\log(I(S_1,S_2;h(X))+1)-4\right)\\&\geq 0.
%\end{align*}
%Moreover, we have
%\begin{align}
%&\bar{L}_{3}^{\epsilon}\!-\!\bar{L}_{4}^{\epsilon}\nonumber\\&=\!\alpha_2\! \left(H(S_1|h(X)\!)\!-\!\log(I(S_1;h(X)\!)\!+\!H(S_2|S_1\!)\!+\!1)\!-\!4\right)\nonumber\\& \stackrel{(a)}{\geq} \alpha_2\left(H(S_1|h(X))-I(S_1;h(X))-H(S_2|S_1)-4\right)\nonumber\\&  \stackrel{(b)}{\geq} \alpha_2\left(H(S_1|h(X))-I(S_1;h(X))-H(h(X)|S_1)-4\right)\nonumber\\ &= \alpha_2\left( H(S_1|h(X))-H(h(X))-4\right)\geq 0,\label{jj}
%\end{align} 
%where (a) follows since $\log(1+x)\leq x$ and (b) holds since we have $H(S_2|S_1)\leq H(h(X)|S_1)$ and $H(S_2|h(X))=0$. 
%Furthermore,
%\begin{align}
%&\bar{L}_{3}^{\epsilon}\!-\!L_{1}^{\epsilon}\nonumber\\&\!=\! H(S_1|h(X))\!-\!\log(I(S_1;h(X))\!+\!H(S_2|S_1)\!+\!1)\!-\!4\nonumber\\& \geq H(S_1|h(X))-I(S_1;h(X))-H(S_2|S_1)-4\nonumber\\& \geq H(S_1|h(X))-I(S_1;h(X))-H(h(X)|S_1)-4\nonumber\\ &=  H(S_1|h(X))-H(h(X))-4\geq 0.\label{jjj}
%\end{align}
%Finally, by using \eqref{jj} and \eqref{jjj} we conclude that $L_{3}^{\epsilon}$ is dominant and we have  
\begin{align*}
\bar{L}_{3}^{\epsilon}\geq \max\{L_{2}^{\epsilon},\bar{L}_{4}^{\epsilon},L_{1}^{\epsilon}\}.
\end{align*} 
%\section{Multi-task private semantic communication}
\textbf{Multi-task scenario:}
The next theorem helps us to derive upper bounds on privacy-utility trade-off in \eqref{main1}.
\begin{theorem}\label{theorem1}
	For any feasible $U$ in \eqref{main1}, there exists RV $U^*=(U^*_1,...,U^*_N)$ with conditional distribution $P_{U^*|XS}(u^*|x,s)=\prod_{i=1}^NP_{U^*_i|X_iS_i}(u^*_i|x_i,s_i)$ that obtains the same leakage as $U$, i.e, we have
	\vspace{-2mm}
	\begin{align}\label{koon}
	I(S;U^*)=I(S;U),
	\end{align} 
	and for all $i\in\{1,...,L\}$ it bounds the utility achieved by $U$ as follows
	\begin{align}\label{koonkesh}
	I(h_i(X);U)\leq I(h_i(X);U^*)+\min\{\Delta_i^1,\Delta_i^2\}, 
	\end{align}
	where
	%\begin{align*}
	$\Delta_i^1=I(S;h_i(X))+\sum_{j:X_i\in h_j(X)}H(S_j|X_j)=\sum_{j:X_i\in h_j(X)} \left( I(S_j;X_j)+H(S_j|X_j)\right)$ and
	$\Delta_i^2=I(S;h_i(X))+\sum_{j:X_i\in h_j(X)}\left(\log(I(S_j;X_j)+1)+4\right)=\sum_{j:X_i\in h_j(X)} \left( I(S_j;X_j)+\log(I(S_j;X_j)+1)+4\right)$.
	%	\Delta_i=\sum_{j:Y_j\notin C_i} H(X_j)\min\{\!\!\!\sum_{j:Y_i\in C_j}\!\!\!\!\left(\log(I(X_j;Y_i)+1)+4\right)\!,\!\!\sum_{j:Y_i\in C_j}\!\!\!H(X_j|Y_j)\}.
	%\end{align*}
\end{theorem}
\textbf{Proof:} The proof is provided in Appendix C.
\begin{theorem}\label{theorem2}
	For any $(S,X)$ distributed according $P_{SX}(s,x)=\prod_{i=1}^N P_{S_i,X_i}(s_i,x_i)$ we have
	\begin{align}\label{jaleb}
	&h_{\epsilon}^L(P_{S,f(X),\bm{h}(X)})\!\leq\nonumber\\ &\epsilon\max_i\left\{\sum_{j:X_i\in h_j}\!\!\!\lambda_j\right\}\!+\!\sum_{i=1}^N\!\left(\sum_{j:X_i\in h_j}\!\!\!\lambda_j\right)\!\!\left(H(X_i|S_i)\!+\!\delta_i\right),
	\end{align}
	where $\delta_i=\min\{\Delta_i^1,\Delta_i^2\}$, $\Delta_i^1$ and $\Delta_i^2$ are defined in Theorem~\ref{theorem1}.
\end{theorem}
\textbf{Proof:}
	By using Theorem~\ref{theorem1} we decompose $I(U;h_i(X))$ and $I(U;S)$ into $N$ parts as follows
	\begin{align}
	&h_{\epsilon}^L\leq \!\!\!\!\!\!\!\sup_{\begin{array}{c}\substack{\{P_{U_i|S_i,X_i}\}_{i=1}^N:\\\sum_{i=1}^NI(S_i;U_i)\leq \epsilon}\end{array}}\!\!\sum_{i=1}^N\!\left(\sum_{j:X_i\in h_j}\!\!\!\lambda_j\right)\!\!\left(I(U_i;X_i)\!+\!\delta_i\right)\label{ame1}\\&=\!\!\!\!\!\!\!\sup_{\begin{array}{c}\substack{\{P_{U_i|S_i,X_i},\epsilon_i\}_{i=1}^N:\\I(S_i;U_i)\leq\epsilon_i,\\ \sum_i \epsilon_i\leq \epsilon}\end{array}}\!\!\sum_{i=1}^N\!\left(\sum_{j:X_i\in h_j}\!\!\!\lambda_j\right)\!\!\left(I(U_i;X_i)\!+\!\delta_i\right)\label{nane1}.
	\end{align}
	Note that we can maximize \eqref{nane1} in two phases. First, we fix $\epsilon_i$ and decompose \eqref{nane1} into $N$ problems and find an upper bound for each of them. We then show that the upper bound can be written as a linear program over $\{\epsilon_i\}_{i=1}^N$ and we obtain the final upper bound. Thus, by using the upper bound in \cite[Lemma~6]{shah} we have
		\begin{align*}
	&h_{\epsilon}^L\!\leq\!\!\!\!\!\!\! \sup_{\begin{array}{c}\substack{\{\epsilon_i\}_{i=1}^N:\\ 0\leq\epsilon_i\leq \epsilon,\\ \sum_i \epsilon_i\leq \epsilon}\end{array}}\!\!\!\sum_{i=1}^N\!\left(\sum_{j:X_i\in h_j}\!\!\!\lambda_j\right)\!\!\left(H(X_i|S_i)\!+\!\epsilon_i\!+\!\delta_i\right)\\&=\epsilon\max_i\left\{\sum_{j:X_i\in h_j}\!\!\!\lambda_j\right\}\!+\!\sum_{i=1}^N\!\left(\sum_{j:X_i\in h_j}\!\!\!\lambda_j\right)\!\!\left(H(X_i|S_i)\!+\!\delta_i\right).
	\end{align*}
	The last line follows since the terms $\sum_{j:X_i\in h_j}\lambda_j$, $\sum_{i=1}^N\sum_{j:X_i\in h_j}\lambda_j\delta_i$ and $\sum_{i=1}^N\sum_{j:X_i\in h_j}\lambda_jH(S_i|X_i)$ are constant, thus, we choose $\epsilon_i=\epsilon$ with the largest coefficient. \\
Next we find lower bounds on $h_{\epsilon}^L(P_{S,f(X),\bm{h}(X)})$.
\begin{theorem}\label{theorem3}
	Let $\gamma_i=1-\frac{H(S_i|X_i)}{H(S_i)}+\frac{\log(I(S_i;X_i)+1)+4}{H(S_i)}$ and $\mu_i=\sum_{j:X_i\in h_j}\lambda_j$. For any $(S,X)$ distributed according $P_{SX}(s,x)=\prod_{i=1}^N P_{S_i,X_i}(s_i,x_i)$ we have
	\begin{align*}
	h_{\epsilon}^L(P_{S,f(X),\bm{h}(X)})\geq \max\{L_{\epsilon}^1,L_{\epsilon}^2,L_{\epsilon}^3,L_{\epsilon}^4\},
	\end{align*}
	where
	\begin{align*}
	L_{\epsilon}^1&=\epsilon\max_i\{\mu_i\}+\sum_{i=1}^N\mu_i\left(H(X_i|S_i)-H(S_i|X_i)\right),\\
	L_{\epsilon}^2&=\sum_{i=1}^N\! \mu_i\left(H(X_i|S_i)-(\log(I(S_i;X_i)+1)+4)\right)\\&+\max_i\{\mu_i\gamma_i\}\epsilon,\\
	L_{\epsilon}^3&=\epsilon\max_i\{\mu_i(1-\min_{(S_{i_1},S_{i_2})\in\mathcal{K}_{S_i}}\frac{H(S_{i_2}|X_i)}{H(S_{i_2})})\}\\&+\sum_{i=1}^N\! \mu_i\left(H(X_i|S_i)-(\log(I(S_i;X_i)+1)+4)\right)\\
	L_{\epsilon}^4&=\sum_{i=1}^N\! \mu_i\left(H(X_i|S_i)-(\log(I(S_i;X_i)+1)+4)\right)+ \\& \!\!\!\!\!\epsilon\max_i\{\mu_i(1\!-\!\!\!\!\!\!\!\!\min_{(S_{i_1},S_{i_2})\in\mathcal{K}_{S_i}}\!\!\!\!\!\!\!\!\!\!\!\frac{H(S_{i}|X_i)\!-\!(\log(I(S_i;X_i)\!+\!1)\!+\!4)}{H(S_{i_2})})\}
	\end{align*}
	%$
	%L_{\epsilon}^1=\epsilon\max_i\{\mu_i\}+\sum_{i=1}^N\mu_i\left(H(X_i|S_i)-H(S_i|X_i)\right),$ and $
	%L_{\epsilon}^2=\sum_{i=1}^N\! \mu_i\left(H(X_i|S_i)-(\log(I(S_i;X_i)+1)+4)\right)+\max_i\{\mu_i\gamma_i\}\epsilon,
	%$ $L_{\epsilon}^3=\epsilon\max_i\{\mu_i(1-\min_{(S_{i_1},S_{i_2})\in\mathcal{K}_{S_i}}\frac{H(S_{i_2}|X_i)}{H(S_{i_2})})\}+\sum_{i=1}^N\! \mu_i\left(H(X_i|S_i)-(\log(I(S_i;X_i)+1)+4)\right)$, and $L_{\epsilon}^4=\epsilon\max_i\{\mu_i(1-\min_{(S_{i_1},S_{i_2})\in\mathcal{K}_{S_i}}\frac{H(S_{i}|X_i)-(\log(I(S_i;X_i)+1)+4)}{H(S_{i_2})})\} +\sum_{i=1}^N\! \mu_i\left(H(X_i|S_i)-(\log(I(S_i;X_i)+1)+4)\right)$. 
	Moreover $\mathcal{K}_{S_i}$ is the set of all representations of $S_i$ using separation technique.
\end{theorem}
\textbf{Proof:}
	To find the first lower bound, let us construct $\tilde{U}=(\tilde{U}_1,..,\tilde{U}_N)$ as follows: 
	Let $\tilde{U}_i$ be the RV found by the EFRL in \cite[Lemma~4]{shah} using $X\leftarrow S_i$, $Y\leftarrow X_i$ and $\epsilon\leftarrow \epsilon_i$. The utility attained by $\tilde{U}_i$ can be lower bounded as follows
	\begin{align}\label{ja}
	&I(\tilde{U}_i;X_i)\nonumber\\
	&=I(\tilde{U}_i;X_i)+H(X_i|S_i)-I(S_i;\tilde{U}_i|X_i)-H(X_i|\tilde{U}_i,S_i)\nonumber\\
	&\geq H(X_i|S_i)-H(S_i|X_i)+\epsilon_i, 
	\end{align}
	 Furthermore, similar to the upper bounds, we can construct $\tilde{U}$ so that $\{(\tilde{U}_i,X_i,S_i)\}_{i=1}^N$ become mutually independent. Thus, we have
	 \begin{align}\label{jaleb1}
	 &h_{\epsilon}^L(P_{S,f(X),\bm{h}(X)})\!\geq\nonumber\\ &\epsilon\max_i\!\left\{\!\sum_{j:Y_i\in h_j}\!\!\!\!\lambda_j\!\right\}\!+\!\sum_{i=1}^N\!\left(\sum_{j:Y_i\in h_j}\!\!\!\!\lambda_j\!\!\right)\!\!\left(H(Y_i|X_i)\!-\!H(X_i|Y_i)\!\right).
	 \end{align}
	 To find the second lower bound, let us construct $U'=(U'_1,..,U'_N)$ as follows:
	  Let $U'_i$ be the RV found by the ESFRL in \cite[Lemma~5]{shah} using $X\leftarrow S_i$, $Y\leftarrow X_i$ and $\epsilon\leftarrow \epsilon_i$. The utility attained by $U'_i$ can be lower bounded as follows
	  \begin{align}\label{jaa}
	  &I(U'_i;X_i)
	  \geq H(X_i|S_i)-\alpha_iH(S_i|X_i)+\epsilon_i\nonumber\\&-(1-\alpha_i)\left(\log(I(S_i;X_i)+1)+4\right), 
	  \end{align}
	  where $\alpha_i=\frac{\epsilon_i}{H(S_i)}$.
	  Furthermore, similar to the upper bounds, we can construct $U'$ so that $\{(U'_i,Y_i,X_i)\}_{i=1}^N$ become mutually independent. Hence, we can find the second lower bound for $h_{\epsilon}^L(P_{XY})$ as follows. Let $\beta_i=H(X_i|S_i)-\alpha_iH(S_i|X_i)+\epsilon_i-(1-\alpha_i)\left(\log(I(S_i;X_i)+1)+4\right)$, $\alpha_i=\frac{\epsilon_i}{H(S_i)}$, $\gamma_i=1-\frac{H(S_i|X_i)}{H(S_i)}+\frac{\log(I(S_i;X_i)+1)+4}{H(S_i)}$, and $\mu_i=\sum_{j:X_i\in h_j}\lambda_j$. We have
	  \begin{align*}
	  &h_{\epsilon}^L(P_{S,f(X),\bm{h}(X)})\!\geq \sup_{\begin{array}{c}\substack{\{\epsilon_i\}_{i=1}^N:\\ 0\leq\epsilon_i\leq \epsilon,\\ \sum_i \epsilon_i\leq \epsilon}\end{array}}\sum_{i=1}^N\left(\sum_{j:X_i\in h_j}\!\!\!\lambda_j\right)\beta_i=\\&\!\!\!\!\sup_{\begin{array}{c}\substack{\{\epsilon_i\}_{i=1}^N:\\ 0\leq\epsilon_i\leq \epsilon,\\ \sum_i \epsilon_i\leq \epsilon}\end{array}}\!\!\!\!\sum_{i=1}^N \mu_i\left(H(X_i|S_i)\!-\!(\log(I(S_i;X_i)\!+\!1)\!+\!4)\right)\!+\!\mu_i\gamma_i\epsilon_i\nonumber\\&=\!\!\sum_{i=1}^N \!\mu_i\!\left(H(X_i|S_i\!)\!-\!(\log(I(S_i;\!X_i\!)\!+\!1)\!+\!4)\!\right)\!+\!\max_i\{\mu_i\gamma_i\}\epsilon .
	  \end{align*}
	  %Note that the right hand side of \eqref{jaleb2} is a linear program which can be rewritten as follows: Let $\gamma_i=1-\frac{H(S_i|X_i)}{H(S_i)}+\frac{\log(I(S_i;X_i)+1)+4}{H(S_i)}$ and $\mu_i=\sum_{j:X_i\in h_j}\lambda_j$. We have
%	  \begin{align*}
%	  &\sum_{i=1}^N\left(\sum_{j:X_i\in h_j}\!\!\!\lambda_j\right)\beta_i\!\\&=\!\sum_{i=1}^N \mu_i\left(H(X_i|S_i)\!-\!(\log(I(S_i;X_i)\!+\!1)\!+\!4)\right)\!+\!\mu_i\gamma_i\epsilon_i
%	  \end{align*} 
%	  Thus,
%	  \begin{align}\label{tala}
%	  &\sup_{\begin{array}{c}\substack{\{\epsilon_i\}_{i=1}^N:\\ 0\leq\epsilon_i\leq \epsilon,\nonumber\\ \sum_i \epsilon_i\leq \epsilon}\end{array}}\sum_{i=1}^N\left(\sum_{j:X_i\in h_j}\!\!\!\lambda_j\right)\beta_i=\\&\sum_{i=1}^N \mu_i\left(H(X_i|S_i)\!-\!(\log(I(S_i;X_i)\!+\!1)\!+\!4)\right)\!+\!\max_i\{\mu_i\gamma_i\}\epsilon,
%	  \end{align}
	  where in the last line we choose $\epsilon_i=\epsilon$ with the largest coefficient. Using similar arguments, lower bounds $L_{h}^{3}(\epsilon)$ and $L_{h}^{4}(\epsilon)$, and separation technique we can achieve $L_{\epsilon}^3$ and $L_{\epsilon}^4$. For more details see supplementary file.
	  \begin{remark}
	  	\normalfont
	  	The lower bounds in Theorem 4 assert that we release the information about only one component in $(S_1,\ldots,S_N)$, with the leakage allowed to be equal to the maximum level, i.e., $\epsilon$. In other words, the disclosed data is correlated to one of the components in $S$ and is independent of the others. The upper bound in Theorem 3 suggests a similar mechanism. 
	  \end{remark}
	  \vspace{-2mm}
%\section{conclusion}\label{concull}
%We have introduced 
%a multi-task semantic communication problem with privacy constraint where considering $L=1$ using extended versions of the FRL, SFRL, and separation technique lower bounds on $h_{\epsilon}(P_{S,f(X),h(X)})$ are obtained. %Furthermore the tightness of the bounds has been investigated and is shown that 
%Moreover, when the private data $S$ is a deterministic function of the goal $h(X)$, the upper bound is achieved. Considering multi-task scenario we have shown that the main complex problem can be decomposed to multiple single-task problems and a simple privacy design is presented. %Moreover, we have studied the bounds considering different scenarios. Finally, the problem is studied in a multi-task scenario.  %Moreover, a necessary condition for an optimizer of $h_{\epsilon}(P_{XY})$ has been obtained. In the case of perfect privacy, new lower and upper bounds are derived using ESFRL and excess functional information.
%\vspace{-2mm}
\section*{Appendix A}
%\subsection*{Proof of observation}
\begin{observ}(Separation technique)
	Any discrete RV $S$ supported on $\mathcal{S}=\{1,\ldots,|\mathcal{S}|\}$ can be represented by two RVs $(S_1,S_2)$ where $1<|\mathcal{S}_1|<|\mathcal{S}|$ and $1<|\mathcal{S}_2|<|\mathcal{S}|$. 
\end{observ}
\textbf{Proof:} The proof is provided in the supplementary material.
%	First, let $|\mathcal{S}|$ be not a prime number. Thus, there exist $|\mathcal{S}_1|$ and $|\mathcal{S}_2|$ such that $|\mathcal{S}|=|\mathcal{S}_1|\times|\mathcal{S}_2|$ where $|\mathcal{S}_1|\geq|\mathcal{S}_2|\geq 2$. We can uniquely map each $x\in\mathcal{S}$ into a pair $(s_1,s_2)$ where $s_1\in\mathcal{S}_1$ and $s_2\in\mathcal{S}_2$. As a result, we can represent $S$ by the pair $(S_1,S_2)$ where $\mathcal{S}_1= \{1,\ldots,|\mathcal{S}_1|\}$, $\mathcal{S}_2= \{1,\ldots,|\mathcal{S}_2|\}$, and $P_{S}(s)=P_{S_1S_2}(s_1,s_2)$. Next, let $|\mathcal{S}|$ be a prime number. Hence, there exist $|\mathcal{S}_1|$ and $|\mathcal{S}_2|$ such that $|\mathcal{S}|+1=|\mathcal{S}_1|\times|\mathcal{S}_2|$ and we can represent $S$ by the pair $(S_1,S_2)$ where $P_{S_1S_2}(s_1=|\mathcal{S}_1|,s_2=|\mathcal{S}_2|)=0$. In other words, the last pair $(|\mathcal{S}_1|,|\mathcal{S}_2|)$ is not mapped to any $s\in\mathcal{S}$. 
\vspace{-2mm}
\section*{Appendix B}
A detailed proof is provided in the supplementary material. Let $\bar{U}$ be found by SFRL with $S=(S_1,S_2)\leftarrow X$ and $h(X)\leftarrow Y$. 
\iffalse
We have
\begin{align*}
I(\bar{U};S_1,S_2)&=H(h(X)|\bar{U},S_1,S_2)=0,\\
I(S_1,S_2;\bar{U}|h(X))&\leq \log(I(S_1,S_2;h(X))+1)+4.
\end{align*}
\fi
Moreover, let $U=(\bar{U},W)$ with $W=\begin{cases}
S_2,\ \text{w.p}.\ \alpha_2\\
c,\ \ \text{w.p.}\ 1-\alpha_2
\end{cases}$, where $c$ is a constant and $\alpha_2=\frac{\epsilon}{H(S_2)}$. We have %First we show that $I(U;S_1,S_2)=\epsilon$. We have
\begin{align*}
I(U\!;\!S_1,S_2)\!=\!I(\bar{U},W\!;\!S_1,S_2)\!=\!I(W\!;\!S_1,S_2)\!=\!\alpha H(\!S_2\!)\!=\!\epsilon,
\end{align*}
%where (a) follows since $\bar{U}$ is independent of $(S_1,S_2,W)$.
Next, we expand $I(U;S_1,S_2|h(X))$.
%\begin{align*}
%I(U;X_1,X_2)&=I(\bar{U},W;X_1,X_2)\stackrel{(a)}{=}I(W;X_1,X_2)\\&=\!H\!(X_1,\!X_2)\!-\!\alpha H(X_1|X_2)\!-\!(1\!-\!\alpha)H\!(X_1,\!X_2)\\&=\alpha H(X_2)=\epsilon,
%\end{align*}
%where (a) follows since $\bar{U}$ is independent of $(X_1,X_2,W)$. Furthermore, we have
\begin{align}
&I(U;S_1,S_2|h(X))\\&=I(\bar{U};S_1,S_2|h(X))+I(W;S_1,S_2|h(X),\bar{U})\nonumber\\&=\!(1\!-\!\alpha_2)I(\bar{U};S_1,\!S_2|h(X))\!+\!\alpha_2 H(S_1,S_2|h(X))\nonumber\\&-\!\alpha_2 H(S_1|h(X),\bar{U},S_2).\label{jakesh}
\end{align}
In the following we bound \eqref{jakesh} in two ways. We have
\begin{align}
&\eqref{jakesh}=I(\bar{U};S_1,\!S_2|h(X))\!+\!\alpha_2 H(S_2|h(X))\!\\&-\!\alpha_2 I(\bar{U};S_2|h(X))\nonumber\\ &\stackrel{(a)}{\leq} \log(I(S_1,S_2;h(X))+1)\!+\!4\!+\!\alpha_2 H(S_2|h(X)).\label{jakesh2}
\end{align}
Furthermore,
\begin{align}
&\eqref{jakesh}\leq \!(1\!-\!\alpha_2)I(\bar{U};S_1,\!S_2|h(X))+\alpha_2 H(S_1,\!S_2|h(X)\!) \nonumber\\&\stackrel{(b)}{\leq}\! \!(\!1\!\!-\!\!\alpha_2)\!\left(\log(I(S_1,\!S_2;\!h(X)\!)\!+\!1)\!+\!4\right)\!+\!\alpha_2 H(\!S_1,\!S_2|h(X)\!).\label{jakesh3}
\end{align}
Inequalities (a) and (b) follow since $\bar{U}$ is produced by SFRL. %so that $I(\bar{U};S_1,S_2|h(X))\leq \log(I(S_1,S_2;h(X))+1)+4$. 
Using \eqref{jakesh2}, \eqref{jakesh3}, and \eqref{key} we have
\begin{align}
h_{\epsilon}&\geq \epsilon+H(h(X)|S)-\alpha_2 H(S_2|h(X))\nonumber\\&-\left(\log(I(S_1,S_2;h(X))+1)+4\right),\label{as}\\
h_{\epsilon}&\geq \epsilon+H(h(X)|S)-\alpha_2 H(S|h(X))\nonumber\\&-(1-\alpha_2)(\log(I(S;h(X))+1)+4).\label{ass}
\end{align} 
%In steps (c) and (d) we used $H(h(X)|S_1,S_2,U)=0$. The latter follows by definition of $W$ and the fact that $\bar{U}$ is produced by SFRL. 
Note that since both \eqref{as} and \eqref{ass} hold for any representation of $S$ we can take maximum over all possible representations and we obtain the final results.
%\begin{align*}
%&h_{\epsilon}(P_{S,f(X),h(X)})\\&\geq  H(h(X)|S)+\epsilon-\left( \log(I(S;h(X))+1)+4 \right)\\&-\min_{(S_1,S_2)\in \mathcal{K}_S} \{\alpha_2 H(S_2|Y)\}=L_{h}^{3}(\epsilon),\\
%&h_{\epsilon}(P_{S,f(X),h(X)})\\&\geq H(h(X)|S)+\epsilon-\left( \log(I(S;h(X))\!+\!1)\!+\!4 \right)\\&-\!\!\!\min_{(S_1,S_2)\in \mathcal{K}_S}\!\!\!\{\alpha_2 \left(H(S|h(X))-\log(I(S;h(X))\!+\!1)\!+\!4 )\right) \},\\&=L_{h}^{4}(\epsilon).
%\end{align*}
To design the artificial noise $M$, let RVs $U_1$ and $U_2$ achieve $L_{h}^{3}(\epsilon)$ and $L_{h}^{4}(\epsilon)$. Then, the privacy mechanism design that achieves $L_{h}^{3}(\epsilon)$ and $L_{h}^{4}(\epsilon)$ are obtained by $M_1=U_1-f(X)$ and $M_2=U_2-f(X)$. Finally, the results about tightness can be proved by using \cite[Theorem 2]{shah}.
\vspace{-2mm}
\section*{Appendix C}
A detailed proof is provided in the supplementary material. For any any feasible $U$ in \eqref{main1}, let $\bar{U}=(\bar{U}_1,..,\bar{U}_N)$ be the RV constructed as in \cite[Eq. (25) and Eq. (26)]{liu2020robust} substituting $U\leftarrow Y$, and $\bar{U}\leftarrow U$, and $\tilde{U}=(\tilde{U}_1,..,\tilde{U}_N)$ be constructed as follows: For all $i\in\{1,...,N\}$, let $\tilde{U}_i$ be the RV found by the FRL using $X\leftarrow(\bar{U}_i,X_i)$ and $Y\leftarrow Y_i$. Thus, using proof of \cite[Theorem 1]{liu2020robust} we have
\begin{itemize}
	\item [i.] $\bar{U}-S-(X,U)$ forms a Markov chain.
	\item [ii.] $\{(\bar{U}_i,X_i,S_i)\}_{i=1}^N$ are mutually independent.
	\item [iii.] $I(S;\bar{U})=I(S;U)$.
\end{itemize}
By checking the proof in \cite[Th.~1]{liu2020robust}, the assumption that $S$ is an element-wise deterministic function of $X$ has not been used. Thus, the same proof works for (i) and (iii). For proving (ii) note that by assumption we have $P_{SX}(s,x)=\prod_{i=1}^N P_{S_iX_i}(s_i,x_i)$ so that the same proof as \cite[Th.~1]{liu2020robust} works.  
%Moreover, we have 
%\begin{align}\label{jj}
%I(\tilde{U}_i;\bar{U}_i,S_i)&=0,
%\end{align}
%and
%\begin{align}\label{ii}
%H(X_i|S_i,\bar{U}_i,\tilde{U}_i)&=0.
%\end{align}
\iffalse
Furthermore, for all $i\in\{1,...,N\}$, let $U'_i$ be the RV found by SFRL in Lemma~\ref{lemma2} using $X\leftarrow(\bar{U}_i,X_i)$ and $Y\leftarrow Y_i$. Thus, we have
\begin{align*}
I(\tilde{U}_i;\bar{U}_i,X_i)&=0,
\end{align*}
and
\begin{align*}
H(Y_i|X_i,\bar{U}_i,\tilde{U}_i)&=0,
\end{align*}
and
\begin{align*}
I()\leq
\end{align*}
\fi
Let $U^*=(U^*_1,...,U^*_N)$ where $U^*_i=(\tilde{U}_i,\bar{U}_i)$. Using (ii) and the construction used as in proof of \cite[Lemma~1]{kostala} we can build $\tilde{U}_i$ such that $\{(\tilde{U}_i,\bar{U}_i,S_i,X_i)\}_{i=1}^N$ are mutually independent. Thus, we have %Due to the independence of $(\tilde{U}_i,\bar{U}_i,X_i)$ the conditional distribution $P_{U^*|Y}(u^*|y)$ can be stated as follows
\vspace{-1mm}
\begin{align}
P_{U^*|X,S}(u^*|x,s)\!=\!\prod_{i=1}^{N}P_{U^*|X,S}(u^*_i|s_i,x_i).
\end{align}
For proving the leakage constraint in \eqref{koon} we have
\vspace{-1mm}
\begin{align*}
I(S;U^*)& \stackrel{(a)}{=} \sum_i I(S_i;\tilde{U}_i,\bar{U}_i)\stackrel{(b)}{=}\sum_i I(S_i;\bar{U}_i) = I(S;\bar{U}) \\&\stackrel{(c)}{=} I(S;U),  
\end{align*}
where (a) follows by the independency of $(\tilde{U}_i,\bar{U}_i,Y_i)$ over different $i$'s, (b) follows by $I(\tilde{U}_i;\bar{U}_i,S_i)=0$ and (c) holds due to \cite[Theorem 1]{liu2020robust}. 
For proving %the upper bound on utility in 
\eqref{koonkesh} we first derive expressions for $I(h_i;U^*)$ and $I(h_i;U)$ as follows.
\begin{align}
I(h_i;U^*)
\!&=\!I(S;U^*)\!\!+\!\!H(h_i|X)\!\!-\!\!H(h_i|S,U^*)\!\!-\!\!I(S;U^*|h_i\!).\label{kosenanat}\\
%\end{align} 
%Similarly, we have
%\begin{align}
I(h_i;U)&\!=\!I(S;U)\!+\!H(h_i|S)\!\!-\!\!H(h_i|S,U)\!-\!I(S;U|h_i\!).\label{kosebabat}
\end{align}
\vspace{-1mm}
Thus, by using \eqref{kosenanat}, \eqref{kosebabat} and \eqref{koon} we have
\begin{align*}
&I(h_i;U)
\leq I(h_i;U^*)+\sum_{j:X_i\in h_j} H(X_j|S_j,\bar{U}_j,\tilde{U}_j)\\&+\!\!\sum_{j:X_i\in h_j}\!\! I(S_j;\bar{U}_j,\tilde{U}_j|Y_j)\!+\!\!\!\!\sum_{j:X_i\notin h_j}\!\!\!\! I(S_j;\bar{U}_j,\tilde{U}_j) \!-\!I(S;U|h_i)\\&\stackrel{(a)}{\leq}I(h_i;U^*)+\!\!\sum_{j:X_i\in h_j}\!\! I(S_j;\bar{U}_j,\tilde{U}_j|X_j)\!\\& \ \ +\!\!\!\!\sum_{j:X_i\notin h_j}\!\! I(S_j;\bar{U}_j)-I(S;U|h_i)%&= I(C_i;U^*)\!+\!\!\!\!\!\!\sum_{j:Y_i\in C_j}\!\!\!\!\!\!\left( I(X_j;\bar{U}_j|Y_j)\!+\!I(X_j;\tilde{U}_j|Y_j,\bar{U}_j)\right)\\& \ \ +\!\!\!\!\sum_{j:Y_i\notin C_j}\!\! I(X_j;\bar{U}_j)-I(X;U|C_i)\\      
%&\stackrel{(d)}{\leq} I(h_i;U^*) +\!\!\sum_{j:X_i\in h_j}\!\! H(S_j|X_j)\!+ I(S;U)\\& \ \ -I(S;U|h_i)
%\\&=I(h_i;U^*)\!+\!\!\sum_{j:X_i\in h_j}\!\! H(S_j|X_j)\!+I(S;h_i)\\& \ \ -I(S;h_i|U)
\leq I(h_i;U^*)\\&+\!\!\!\sum_{j:X_i\in h_j}\!\! H(S_j|X_j)\!+I(S;h_i)=I(h_i;U^*)+\Delta_i^1.
\end{align*}
%where in step (a) we used the equality property of the leakage in \eqref{koon},  where step (b) follows by the independency of $(\tilde{U}_i,\bar{U}_i,X_i)$ over different $i$'s and 
where step (a) follows %by the fact that $\tilde{U}_i$ is produced 
by FRL. %i.e.,  $H(X_i|S_i,\bar{U}_i,\tilde{U}_i)=0$ and $\tilde{U}_i$ is independent of $\bar{U}_i$ and $S_i$. Moreover, in step (d) we used the simple bound $\sum_{j:X_i\notin h_j} I(S_j;\bar{U}_j,\tilde{U}_j|X_j)\leq \sum_{j:X_i\notin h_j}H(S_j|X_j)$ and $\sum_{j:X_i\notin h_j}\!\! I(S_j;\bar{U}_j)\leq I(S;\bar{U})=I(S;U)$. %where the last equality follows by Lemma~\ref{pare2}.\\
%For proving the second upper bound let $U'=(U'_1,..,U'_N)$ be constructed as follows. For all $i\in\{1,...,N\}$, let $U'_i$ be the RV found by SFRL \cite[Th. 1]{kosnane} using $X\leftarrow(\bar{U}_i,X_i)$ and $Y\leftarrow Y_i$. Thus, we have
%\begin{align}\label{jjj}
%I(U'_i;\bar{U}_i,S_i)&=0,
%\end{align}
%and
%\begin{align}\label{iii}
%H(X_i|S_i,\bar{U}_i,U'_i)&=0,
%\end{align}
%and
%\begin{align}\label{k}
%I(\bar{U}_i,S_i;U'_i|X_i)&\leq \log(I(\bar{U}_i,S_i;X_i)+1)+4\nonumber\\&=\log(I(S_i;X_i)+1)+4,
%\end{align}
%where in last line we used the Markov chain stated in part (i). 
Now let $U^*=(U^*_1,...,U^*_N)$ where $U^*_i=(U'_i,\bar{U}_i)$. Similarly, we construct $U'$ such that $\{(U'_i,\bar{U}_i,S_i,X_i)\}_{i=1}^N$ are mutually independent. %The proof of \eqref{koon} does not change and 
We have
\begin{align*}
&I(h_i;U)\leq I(h_i;U^*)+\!\!\sum_{j:X_i\in h_j}\!\! I(S_j;\bar{U}_j,U'_j|X_j)\!\\&+\!\!\!\!\sum_{j:X_i\notin h_j}\!\! I(S_j;\bar{U}_j)-I(S;U|h_i)=I(h_i;U^*)\!\\&+\!\!\!\!\!\!\sum_{j:X_i\in h_j}\!\! I(S_j;U'_j|X_j,\bar{U}_j)\! \ \ +\!\!\!\sum_{j:X_i\in h_j}I(S_j;\bar{U}_j|X_j)\\&+\sum_{j:X_i\notin h_j}\!\!\! I(S_j;\bar{U}_j)\!-\!I(S;U|h_i)\leq  I(h_i;U^*)\\&+\!\!\!\!\sum_{j:X_i\in h_j}\!\!\left(\log(I(S_i;X_i)+1)+4\right)\!\! +\!\!\!\sum_{j:X_i\in h_j}I(S_j;\bar{U}_j|X_j)\\&+\sum_{j:X_i\notin h_j}\!\!\! I(S_j;\bar{U}_j)\!-\!I(S;U|h_i)\\&=I(h_i;U^*)+\sum_{j:X_i\in h_j}\!\!\left(\log(I(S_i;X_i)+1)+4\right)\\&I(S;h_i)-I(S;h_i|U)-\sum_{j:X_i\notin h_j}I(\bar{U}_j;X_j)\\&\leq I(h_i;U^*)+\sum_{j:X_i\in h_j}\!\!\left(\log(I(S_i;X_i)+1)+4\right)\\&\ \ +I(S;h_i)=I(h_i;U^*)+\Delta_i^2.
\end{align*}
\clearpage   
\bibliographystyle{IEEEtran}
{\balance \bibliography{IEEEabrv,IZS}}
\end{document}

\newgeometry{left=0.7in,right=0.7in,top=.5in,bottom=1in}
	\title{Supplementary material for ``Multi-Task Private Semantic Communication''}
\vspace{-5mm}
\author{
		\IEEEauthorblockN{Amirreza Zamani, Sajad Daei, Tobias J. Oechtering, Mikael Skoglund \vspace*{0.5em}
			\IEEEauthorblockA{\\
				Division of Information Science and Engineering, KTH Royal Institute of Technology \\
				%$^\ddagger$Dept. of Electrical and Electronic Engineering, Imperial College London\\
				Email: \protect amizam@kth.se, sajado@kth.se, oech@kth.se, skoglund@kth.se }}%\vspace*{-2.1em}
		}
	\maketitle
	\section{Detailed proofs:}
	\subsection{Proof of separation technique:}
	\begin{proof}
		First, let $|\mathcal{S}|$ be not a prime number. Thus, there exist $|\mathcal{S}_1|$ and $|\mathcal{S}_2|$ such that $|\mathcal{S}|=|\mathcal{S}_1|\times|\mathcal{S}_2|$ where $|\mathcal{S}_1|\geq|\mathcal{S}_2|\geq 2$. We can uniquely map each $s\in\mathcal{S}$ into a pair $(s_1,s_2)$ where $s_1\in\mathcal{S}_1$ and $s_2\in\mathcal{S}_2$. As a result, we can represent $S$ by the pair $(S_1,S_2)$ where $\mathcal{S}_1= \{1,\ldots,|\mathcal{S}_1|\}$, $\mathcal{S}_2= \{1,\ldots,|\mathcal{S}_2|\}$, and $P_{S}(s)=P_{S_1S_2}(s_1,s_2)$. Next, let $|\mathcal{S}|$ be a prime number. Hence, there exist $|\mathcal{S}_1|$ and $|\mathcal{S}_2|$ such that $|\mathcal{S}|+1=|\mathcal{S}_1|\times|\mathcal{S}_2|$ and we can represent $S$ by the pair $(S_1,S_2)$ where $P_{S_1S_2}(s_1=|\mathcal{S}_1|,s_2=|\mathcal{S}_2|)=0$. In other words, the last pair $(|\mathcal{S}_1|,|\mathcal{S}_2|)$ is not mapped to any $s\in\mathcal{S}$.   
	\end{proof}
	\subsection{Details of the scenarios in comparison section}
	\textbf{Scenario 1}: To compare $\bar{L}_{4}^{\epsilon}$ with $L_{2}^{\epsilon}$, let us assume that $H(S_1,S_2|h(X))\leq \log(I(S_1,S_2;h(X))+1)+4$. A simple example can be considering $S_1$ and $S_2$ as binary RVs. In this case we have
\begin{align*}
&\bar{L}_{4}^{\epsilon}-L_{2}^{\epsilon}=\epsilon(\frac{1}{H(S_2)}-\frac{1}{H(S_1,S_2)})\times\\&\left(\log(I(S_1,S_2;h(X))+1)+4-H(S_1,S_2|h(X))\right)\geq 0.
\end{align*}
	\textbf{Scenario 2}: To compare $\bar{L}_{3}^{\epsilon}$ with $L_{2}^{\epsilon}$, let us assume that $S_2$ is a deterministic function of $h(X)$ and $H(S_1|h(X))\geq \log(I(S_1,S_2;h(X))+1)+4$. A simple example is to let $4+H(h(X))\leq H(S_1|h(X))$ which leads to $H(S_1|h(X))\geq \log(I(S_1,S_2;h(X))+1)+4$. In this case we have
	\begin{align*}
	&\bar{L}_{3}^{\epsilon}-L_{2}^{\epsilon}=\\&\frac{\epsilon}{H(S_1,S_2)}\left(H(S_1|h(X))\!-\!\log(I(S_1,S_2;h(X))+1)-4\right)\\&\geq 0.
	\end{align*}
	Moreover, we have
	\begin{align}
	&\bar{L}_{3}^{\epsilon}\!-\!\bar{L}_{4}^{\epsilon}\nonumber\\&=\!\alpha_2\! \left(H(S_1|h(X)\!)\!-\!\log(I(S_1;h(X)\!)\!+\!H(S_2|S_1\!)\!+\!1)\!-\!4\right)\nonumber\\& \stackrel{(a)}{\geq} \alpha_2\left(H(S_1|h(X))-I(S_1;h(X))-H(S_2|S_1)-4\right)\nonumber\\&  \stackrel{(b)}{\geq} \alpha_2\left(H(S_1|h(X))-I(S_1;h(X))-H(h(X)|S_1)-4\right)\nonumber\\ &= \alpha_2\left( H(S_1|h(X))-H(h(X))-4\right)\geq 0,\label{jj}
	\end{align} 
	where (a) follows since $\log(1+x)\leq x$ and (b) holds since we have $H(S_2|S_1)\leq H(h(X)|S_1)$ and $H(S_2|h(X))=0$. 
	Furthermore,
	\begin{align}
	&\bar{L}_{3}^{\epsilon}\!-\!L_{1}^{\epsilon}\nonumber\\&\!=\! H(S_1|h(X))\!-\!\log(I(S_1;h(X))\!+\!H(S_2|S_1)\!+\!1)\!-\!4\nonumber\\& \geq H(S_1|h(X))-I(S_1;h(X))-H(S_2|S_1)-4\nonumber\\& \geq H(S_1|h(X))-I(S_1;h(X))-H(h(X)|S_1)-4\nonumber\\ &=  H(S_1|h(X))-H(h(X))-4\geq 0.\label{jjj}
	\end{align}
	Finally, by using \eqref{jj} and \eqref{jjj} we conclude that $L_{3}^{\epsilon}$ is dominant and we have  
	\begin{align*}
	\bar{L}_{3}^{\epsilon}\geq \max\{L_{2}^{\epsilon},\bar{L}_{4}^{\epsilon},L_{1}^{\epsilon}\}.
	\end{align*} 
	\subsection{Detailed proof of Theorem 1:}
	\textbf{Appendix B:}	Let $\bar{U}$ be found by SFRL with $S=(X_1,X_2)\leftarrow X$ and $h(X)\leftarrow Y$. 
	We have
	\begin{align*}
	I(\bar{U};S_1,S_2)&=H(h(X)|\bar{U},S_1,S_2)=0,\\
	I(S_1,S_2;\bar{U}|h(X))&\leq \log(I(S_1,S_2;h(X))+1)+4.
	\end{align*}
	Moreover, let $U=(\bar{U},W)$ with $W=\begin{cases}
	S_2,\ \text{w.p}.\ \alpha_2\\
	c,\ \ \text{w.p.}\ 1-\alpha_2
	\end{cases}$, where $c$ is a constant which does not belong to $\mathcal{S}_1\cup \mathcal{S}_2 \cup \mathcal{X}$ and $\alpha_2=\frac{\epsilon}{H(S_2)}$. First we show that $I(U;S_1,S_2)=\epsilon$. We have
	\begin{align*}
	I(U;S_1,S_2)&=I(\bar{U},W;S_1,S_2)\stackrel{(a)}{=}I(W;S_1,S_2)\\&=\!H\!(S_1,\!S_2)\!-\!\alpha_2 H(S_1|S_2)\!\\&-\!(1\!-\!\alpha_2)H\!(S_1,\!S_2)\\&=\alpha H(S_2)=\epsilon,
	\end{align*}
	where (a) follows since $\bar{U}$ is independent of $(S_1,S_2,W)$.
	Next, we expand $I(U;S_1,S_2|h(X))$.
	%\begin{align*}
	%I(U;X_1,X_2)&=I(\bar{U},W;X_1,X_2)\stackrel{(a)}{=}I(W;X_1,X_2)\\&=\!H\!(X_1,\!X_2)\!-\!\alpha H(X_1|X_2)\!-\!(1\!-\!\alpha)H\!(X_1,\!X_2)\\&=\alpha H(X_2)=\epsilon,
	%\end{align*}
	%where (a) follows since $\bar{U}$ is independent of $(X_1,X_2,W)$. Furthermore, we have
	\begin{align}
	&I(U;S_1,S_2|h(X))\\&=I(\bar{U};S_1,S_2|h(X))+I(W;S_1,S_2|h(X),\bar{U})\nonumber\\&=I(\bar{U};S_1,S_2|h(X))+H(S_1,S_2|h(X),\bar{U})\!\\&-\!H(S_1,S_2|h(X),\bar{U},W)\nonumber\\&=I(\bar{U};S_1,S_2|h(X))+\alpha_2 H(S_1,S_2|h(X),\bar{U})\\&-\alpha_2 H(S_1|h(X),\bar{U},S_2)\nonumber \\&=I(\bar{U};S_1,S_2|h(X))-\alpha_2 H(S_1|h(X),\bar{U},S_2)\\&+\alpha_2\left( H(S_1,S_2|h(X))-I(\bar{U};S_1,S_2|h(X))\right)\nonumber\\&=\!(1\!-\!\alpha_2)I(\bar{U};S_1,\!S_2|h(X))\!+\!\alpha_2 H(S_1,S_2|h(X))\nonumber\\&-\!\alpha_2 H(S_1|h(X),\bar{U},S_2).\label{jakesh}
	\end{align}
	In the following we bound \eqref{jakesh} in two ways. We have
	\begin{align}
	\eqref{jakesh}&=\!(1\!-\!\alpha_2)I(\bar{U};S_1,\!S_2|h(X))\!+\!\alpha_2 H(S_2|h(X))\\&+\alpha I(S_1;\bar{U}|h(X),S_2)\nonumber\\&=I(\bar{U};S_1,\!S_2|h(X))\!+\!\alpha_2 H(S_2|h(X))\!\\&-\!\alpha_2 I(\bar{U};S_2|h(X))\nonumber\\ &\stackrel{(a)}{\leq} \log(I(S_1,S_2;h(X))+1)+4+\alpha_2 H(S_2|h(X)).\label{jakesh2}
	\end{align}
	Furthermore,
	\begin{align}
	\eqref{jakesh}&\leq \!(1\!-\!\alpha_2)I(\bar{U};S_1,\!S_2|h(X))+\alpha_2 H(S_1,\!S_2|h(X)\!) \nonumber\\&\stackrel{(b)}{\leqq} \!(1\!-\!\alpha_2)\left(\log(I(S_1,S_2;h(X))+1)+4\right)\nonumber\\&+\alpha_2 H(S_1,\!S_2|h(X)\!).\label{jakesh3}
	\end{align}
	Inequalities (a) and (b) follow since $\bar{U}$ is produced by SFRL, so that $I(\bar{U};S_1,S_2|h(X))\leq \log(I(S_1,S_2;h(X))+1)+4$. Using \eqref{jakesh2}, \eqref{jakesh3} and key equation in (2) we have
	\begin{align}
	h_{\epsilon}(P_{XY})&\geq I(U;h(X))\nonumber\\&\stackrel{(c)}{\geq} \epsilon+H(h(X)|S_1,S_2)-\alpha_2 H(S_2|h(X))\nonumber\\&-\left(\log(I(S_1,S_2;h(X))+1)+4\right) \nonumber\\&=\epsilon+H(h(X)|S)-\alpha_2 H(S_2|h(X))\nonumber\\&-\left(\log(I(S_1,S_2;h(X))+1)+4\right),\label{as}
	\end{align} 
	and 
	\begin{align}
	h_{\epsilon}(P_{XY})&\geq I(U;h(X))\nonumber\\&\stackrel{(d)}{\geq}  \epsilon+H(h(X)|S_1,S_2)-\alpha_2 H(S_1,S_2|h(X))\nonumber\\&-(1-\alpha_2)(\log(I(S_1,S_2;h(X))+1)+4)\nonumber \\&=\epsilon+H(h(X)|S)-\alpha_2 H(S|h(X))\nonumber\\&-(1-\alpha_2)(\log(I(S;h(X))+1)+4).\label{ass}
	\end{align} 
	In steps (c) and (d) we used $H(h(X)|S_1,S_2,U)=0$. The latter follows by definition of $W$ and the fact that $\bar{U}$ is produced by SFRL. Note that since both \eqref{as} and \eqref{ass} hold for any representation of $X$ we can take maximum over all possible representations and we obtain
	\begin{align*}
	&h_{\epsilon}(P_{S,f(X),h(X)})\\&\geq  H(h(X)|S)+\epsilon-\left( \log(I(S;h(X))+1)+4 \right)\\&-\min_{(S_1,S_2)\in \mathcal{K}_S} \{\alpha_2 H(S_2|Y)\}=L_{h}^{3}(\epsilon),\\
	&h_{\epsilon}(P_{S,f(X),h(X)})\\&\geq H(h(X)|S)+\epsilon-\left( \log(I(S;h(X))\!+\!1)\!+\!4 \right)\\&-\!\!\!\min_{(S_1,S_2)\in \mathcal{K}_S}\!\!\!\{\alpha_2 \left(H(S|h(X))-\log(I(S;h(X))\!+\!1)\!+\!4 )\right) \},\\&=L_{h}^{4}(\epsilon).
	\end{align*}
	To design the artificial noise $M$, let RVs $U_1$ and $U_2$ achieve $L_{h}^{3}(\epsilon)$ and $L_{h}^{4}(\epsilon)$. Then, the privacy mechanism design that achieves $L_{h}^{3}(\epsilon)$ and $L_{h}^{4}(\epsilon)$ are obtained by $M_1=U_1-f(X)$ and $M_2=U_2-f(X)$. Finally, the results about tightness can be proved by using \cite[Theorem 2]{king1}.
	\subsection{Detailed proof of Theorem 2:}
	For any any feasible $U$ in the main problem, let $\bar{U}=(\bar{U}_1,..,\bar{U}_N)$ be the RV constructed as in \cite[Eq. (25) and Eq. (26)]{liu2020robust} substituting $U\leftarrow Y$, and $\bar{U}\leftarrow U$, and $\tilde{U}=(\tilde{U}_1,..,\tilde{U}_N)$ be constructed as follows: For all $i\in\{1,...,N\}$, let $\tilde{U}_i$ be the RV found by the FRL using $X\leftarrow(\bar{U}_i,X_i)$ and $Y\leftarrow Y_i$. Thus, using proof of \cite[Theorem 1]{liu2020robust} we have
	\begin{itemize}
		\item [i.] $\bar{U}-S-(X,U)$ forms a Markov chain.
		\item [ii.] $\{(\bar{U}_i,X_i,S_i)\}_{i=1}^N$ are mutually independent.
		\item [iii.] $I(S;\bar{U})=I(S;U)$.
	\end{itemize}
	By checking the proof in \cite[Th.~1]{liu2020robust}, the assumption that $S$ is an element-wise deterministic function of $X$ has not been used. Thus, the same proof can be used for (i) and (iii). For proving (ii) note that by assumption we have $P_{SX}(s,x)=\prod_{i=1}^N P_{S_iX_i}(s_i,x_i)$ so that the same proof as \cite[Th.~1]{liu2020robust} works.  
	Moreover, we have 
	\begin{align}\label{jj}
	I(\tilde{U}_i;\bar{U}_i,S_i)&=0,
	\end{align}
	and
	\begin{align}\label{ii}
	H(X_i|S_i,\bar{U}_i,\tilde{U}_i)&=0.
	\end{align}
	\iffalse
	Furthermore, for all $i\in\{1,...,N\}$, let $U'_i$ be the RV found by SFRL in Lemma~\ref{lemma2} using $X\leftarrow(\bar{U}_i,X_i)$ and $Y\leftarrow Y_i$. Thus, we have
	\begin{align*}
	I(\tilde{U}_i;\bar{U}_i,X_i)&=0,
	\end{align*}
	and
	\begin{align*}
	H(Y_i|X_i,\bar{U}_i,\tilde{U}_i)&=0,
	\end{align*}
	and
	\begin{align*}
	I()\leq
	\end{align*}
	\fi
	Let $U^*=(U^*_1,...,U^*_N)$ where $U^*_i=(\tilde{U}_i,\bar{U}_i)$. Using (ii) and the construction used as in proof of \cite[Lemma~1]{kostala} we can build $\tilde{U}_i$ such that $\{(\tilde{U}_i,\bar{U}_i,S_i,X_i)\}_{i=1}^N$ are mutually independent. Thus, we have %Due to the independence of $(\tilde{U}_i,\bar{U}_i,X_i)$ the conditional distribution $P_{U^*|Y}(u^*|y)$ can be stated as follows
	\begin{align}
	P_{U^*|X,S}(u^*|x,s)\!=\!\prod_{i=1}^{N}P_{U^*|X,S}(u^*_i|s_i,x_i).
	\end{align}
	For proving the leakage constraint in we have
	\begin{align*}
	I(S;U^*)&= I(S_1,..,S_N;\tilde{U}_1,\bar{U}_1,...,\tilde{U}_N,\bar{U}_N)\\&\stackrel{(a)}{=} \sum_i I(S_i;\tilde{U}_i,\bar{U}_i)\\&\stackrel{(b)}{=}\sum_i I(S_i;\bar{U}_i) = I(S;\bar{U}) \\&\stackrel{(c)}{=} I(S;U),  
	\end{align*}
	where (a) follows by the independency of $(\tilde{U}_i,\bar{U}_i,Y_i)$ over different $i$'s, (b) follows by \eqref{jj} and (c) holds due to \cite[Theorem 1]{liu2020robust}. 
	For proving the upper bound on utility we first derive expressions for $I(h_i;U^*)$ and $I(h_i;U)$ as follows.
	\begin{align}
	I(h_i;U^*)
	\!=\!I(S;U^*)\!\!+\!\!H(h_i|X)\!\!-\!\!H(h_i|S,U^*)\!\!-\!\!I(S;U^*|h_i\!).\label{kosenanat}
	\end{align} 
	Similarly, we have
	\begin{align}
	I(h_i;U)=\!I(S;U)\!\!+\!\!H(h_i|S)\!\!-\!\!H(h_i|S,U)\!\!-\!\!I(S;U|h_i\!).\label{kosebabat}
	\end{align}
	Thus, by using \eqref{kosenanat}, \eqref{kosebabat} and the leakage constraint we have
	\begin{align*}
	I(h_i;U)&\stackrel{(a)}{=}I(h_i;U^*)+H(h_i|S,U^*)\\& \ \ +I(S;U^*|h_i)-H(h_i|S,U)-I(S;U|h_i)\\
	&\stackrel{(b)}{=}I(h_i;U^*)+\sum_{j:X_i\in h_j} H(X_i|S_i,U^*_i)\\& \ \ +\sum_{j:X_i\in h_j} I(S_j;U^*_j|X_j)+\sum_{j:X_i\notin h_j} I(S_j;U^*_j)\\ \ \ &-H(h_i|S,U)-I(S;U|h_i)\\&\leq I(h_i;U^*)+\sum_{j:X_i\in h_j} H(X_j|S_j,U^*_j)\\& \ \ +\sum_{j:X_i\in h_j} I(S_j;U^*_j|X_j)+\sum_{j:X_i\notin h_j} I(S_j;U^*_j)\\& \ \ -I(S;U|h_i) \\&= I(h_i;U^*)+\sum_{j:X_i\in h_j} H(X_j|S_j,\bar{U}_j,\tilde{U}_j)\\& \ \ +\!\!\sum_{j:X_i\in h_j}\!\! I(S_j;\bar{U}_j,\tilde{U}_j|Y_j)\!+\!\!\!\!\sum_{j:X_i\notin h_j}\!\! I(S_j;\bar{U}_j,\tilde{U}_j)\\& \ \ -I(S;U|h_i)\\&\stackrel{(c)}{=}I(h_i;U^*)+\!\!\sum_{j:X_i\in h_j}\!\! I(S_j;\bar{U}_j,\tilde{U}_j|X_j)\!\\& \ \ +\!\!\!\!\sum_{j:X_i\notin h_j}\!\! I(S_j;\bar{U}_j)-I(S;U|h_i)\\%&= I(C_i;U^*)\!+\!\!\!\!\!\!\sum_{j:Y_i\in C_j}\!\!\!\!\!\!\left( I(X_j;\bar{U}_j|Y_j)\!+\!I(X_j;\tilde{U}_j|Y_j,\bar{U}_j)\right)\\& \ \ +\!\!\!\!\sum_{j:Y_i\notin C_j}\!\! I(X_j;\bar{U}_j)-I(X;U|C_i)\\      
	&\stackrel{(d)}{\leq} I(h_i;U^*) +\!\!\sum_{j:X_i\in h_j}\!\! H(S_j|X_j)\!+ I(S;U)\\& \ \ -I(S;U|h_i)
	\\&=I(h_i;U^*)\!+\!\!\sum_{j:X_i\in h_j}\!\! H(S_j|X_j)\!+I(S;h_i)\\& \ \ -I(S;h_i|U)
	\\&\leq I(h_i;U^*)+\!\!\!\sum_{j:X_i\in h_j}\!\! H(S_j|X_j)\!+I(S;h_i)\\&=I(h_i;U^*)+\Delta_i^1.
	\end{align*}
	where in step (a) we used the equality property of the leakage constraint,  where step (b) follows by the independency of $(\tilde{U}_i,\bar{U}_i,X_i)$ over different $i$'s and step (c) follows by the fact that $\tilde{U}_i$ is produced by FRL, i.e.,  $H(X_i|S_i,\bar{U}_i,\tilde{U}_i)=0$ and $\tilde{U}_i$ is independent of $\bar{U}_i$ and $S_i$. Moreover, in step (d) we used the simple bound $\sum_{j:X_i\notin h_j} I(S_j;\bar{U}_j,\tilde{U}_j|X_j)\leq \sum_{j:X_i\notin h_j}H(S_j|X_j)$ and $\sum_{j:X_i\notin h_j}\!\! I(S_j;\bar{U}_j)\leq I(S;\bar{U})=I(S;U)$. %where the last equality follows by Lemma~\ref{pare2}.\\
	For proving the second upper bound let $U'=(U'_1,..,U'_N)$ be constructed as follows. For all $i\in\{1,...,N\}$, let $U'_i$ be the RV found by SFRL \cite[Th. 1]{kosnane} using $X\leftarrow(\bar{U}_i,X_i)$ and $Y\leftarrow Y_i$. Thus, we have
	\begin{align}\label{jjj}
	I(U'_i;\bar{U}_i,S_i)&=0,
	\end{align}
	and
	\begin{align}\label{iii}
	H(X_i|S_i,\bar{U}_i,U'_i)&=0,
	\end{align}
	and
	\begin{align}\label{k}
	I(\bar{U}_i,S_i;U'_i|X_i)&\leq \log(I(\bar{U}_i,S_i;X_i)+1)+4\nonumber\\&=\log(I(S_i;X_i)+1)+4,
	\end{align}
	where in last line we used the Markov chain stated in part (i). Now let $U^*=(U^*_1,...,U^*_N)$ where $U^*_i=(U'_i,\bar{U}_i)$. Similarly, we can construct $U'$ such that $\{(U'_i,\bar{U}_i,S_i,X_i)\}_{i=1}^N$ are mutually independent. The proof of the leakage constraint does not change and for proving the second upper bound on utility we have
	\begin{align*}
	I(h_i;U)&\leq I(h_i;U^*)+\!\!\sum_{j:X_i\in h_j}\!\! I(S_j;\bar{U}_j,U'_j|X_j)\!\\& \ \ +\!\!\!\!\sum_{j:X_i\notin h_j}\!\! I(S_j;\bar{U}_j)-I(S;U|h_i)\\&= I(h_i;U^*)\!+\!\!\!\!\!\!\sum_{j:X_i\in h_j}\!\! I(S_j;U'_j|X_j,\bar{U}_j)\!\\& \ \ +\!\!\!\sum_{j:X_i\in h_j}I(S_j;\bar{U}_j|X_j)+\sum_{j:X_i\notin h_j}\!\!\! I(S_j;\bar{U}_j)\!\\&-\!I(S;U|h_i)\\&\stackrel{(a)}{\leq}  I(h_i;U^*)+\sum_{j:X_i\in h_j}\!\!\left(\log(I(S_i;X_i)+1)+4\right)\\&\ \ +\!\!\!\sum_{j:X_i\in h_j}I(S_j;\bar{U}_j|X_j)+\sum_{j:X_i\notin h_j}\!\!\! I(S_j;\bar{U}_j)\!\\&-\!I(S;U|h_i)\\&\stackrel{(b)}{=}I(h_i;U^*)+\sum_{j:X_i\in h_j}\!\!\left(\log(I(S_i;X_i)+1)+4\right)\\&I(S;h_i)-I(S;h_i|U)-\sum_{j:X_i\notin h_j}I(\bar{U}_j;X_j)\\&\leq I(h_i;U^*)+\sum_{j:X_i\in h_j}\!\!\left(\log(I(S_i;X_i)+1)+4\right)\\&\ \ +I(S;h_i)=I(h_i;U^*)+\Delta_i^2.
	\end{align*}
	Step (a) follows by \eqref{k} since we have
	\begin{align*}
	I(S_i;U'_i|X_i,\bar{U}_i)\leq I(\bar{U}_i,S_i;U'_i|X_i).
	\end{align*}
	Furthermore, step (b) follows since we have
	\begin{align*}
	&\sum_{j:X_i\in h_j}\!\!I(S_j;\bar{U}_j|X_j)+\!\!\sum_{j:X_i\notin h_j}\!\!\! I(S_j;\bar{U}_j)-I(S;U|h_i)\stackrel{(a)}{=}\\&\sum_{j:X_i\in h_j}\!\! H(\bar{U}_j|X_j)\!\!+\!\!\sum_{j:X_i\notin h_j} \!\!H(\bar{U}_j)\!-\!H(\bar{U}|S)\!-\!I(S;U|h_i)=\\&\sum_{j:X_i\in h_j}\!\! H(\bar{U}_j|X_j)+\sum_{j:X_i\notin h_j}\!\! \!\!H(\bar{U}_j)\!-\!H(\bar{U})\!\\&\ \ I(\bar{U};S)\!-\!I(S;U|h_i)=\\& \ \ -\sum_{j:X_i\in h_j} I(\bar{U}_j;X_j)+I(S;U)-I(S;U|h_i)=\\&\ \ -\sum_{j:X_i\in h_j} I(\bar{U}_j;X_j)+I(S;h_i)-I(S;h_i|U),
	\end{align*}
	where step (a) follows by the Markov chain $\bar{U}-S-X$.
	\subsection{Details of Theorem 4: acheivability of $L_{\epsilon}^3$ and $L_{\epsilon}^4$}
	To attain $L_{\epsilon}^3$ let  $\tilde{U}=(\tilde{U}_1,..,\tilde{U}_N)$ and $\tilde{U}_i$ be the RV that achieves $L_{h}^{3}(\epsilon)$. Similarly, we can construct $\tilde{U}$ so that $\{(\tilde{U}_i,X_i,S_i)\}_{i=1}^N$ become mutually independent. 
	Let $\beta_i=H(X_i|S_i)\!-\!(\log(I(S_i;X_i)\!+\!1)\!+\!4)+\epsilon_i-\min_{(S_{i_1},S_{i_2})\in\mathcal{K}_{S_i}}\frac{H(S_{i_2}|X_i)}{H(S_{i_2})}$, $\mathcal{K}_{S_i}$ be the set of all representations of $S_i$ using separation technique, and $\gamma_i=1-\min_{(S_{i_1},S_{i_2})\in\mathcal{K}_{S_i}}\frac{H(S_{i_2}|X_i)}{H(S_{i_2})}$
	We have
 \begin{align*}
&h_{\epsilon}^L(P_{S,f(X),\bm{h}(X)})\!\geq \sup_{\begin{array}{c}\substack{\{\epsilon_i\}_{i=1}^N:\\ 0\leq\epsilon_i\leq \epsilon,\\ \sum_i \epsilon_i\leq \epsilon}\end{array}}\sum_{i=1}^N\left(\sum_{j:X_i\in h_j}\!\!\!\lambda_j\right)\beta_i=\\&\!\!\!\!\sup_{\begin{array}{c}\substack{\{\epsilon_i\}_{i=1}^N:\\ 0\leq\epsilon_i\leq \epsilon,\\ \sum_i \epsilon_i\leq \epsilon}\end{array}}\!\!\!\!\sum_{i=1}^N \mu_i\left(H(X_i|S_i)\!-\!(\log(I(S_i;X_i)\!+\!1)\!+\!4)\right)\!+\!\mu_i\gamma_i\epsilon_i\nonumber\\&=\!\!\sum_{i=1}^N \!\mu_i\!\left(H(X_i|S_i\!)\!-\!(\log(I(S_i;\!X_i\!)\!+\!1)\!+\!4)\!\right)\!+\!\max_i\{\mu_i\gamma_i\}\epsilon .
\end{align*}
	Finally, to attain $L_{\epsilon}^4$ let $\gamma_i=1-\min_{(S_{i_1},S_{i_2})\in\mathcal{K}_{S_i}}\frac{H(S_{i}|X_i)-(\log(I(S_i;X_i)+1)+4)}{H(S_{i_2})}$. We have
	\begin{align*}
	&h_{\epsilon}^L(P_{S,f(X),\bm{h}(X)})\!\geq \sup_{\begin{array}{c}\substack{\{\epsilon_i\}_{i=1}^N:\\ 0\leq\epsilon_i\leq \epsilon,\\ \sum_i \epsilon_i\leq \epsilon}\end{array}}\sum_{i=1}^N\left(\sum_{j:X_i\in h_j}\!\!\!\lambda_j\right)\beta_i\\&=\!\!\sum_{i=1}^N \!\mu_i\!\left(H(X_i|S_i\!)\!-\!(\log(I(S_i;\!X_i\!)\!+\!1)\!+\!4)\!\right)\!+\!\max_i\{\mu_i\gamma_i\}\epsilon .
	\end{align*}
	\bibliographystyle{IEEEtran}
	\bibliography{IEEEabrv,IZS}